\documentclass[usenatbib,fleqn]{mn2e}
\usepackage{fixltx2e,longtable,ifpdf,amsmath,amssymb}
\usepackage{times}
\ifpdf
  \usepackage{aeguill}
  \usepackage[pdftex]{graphicx,color}
\else
  \usepackage[dvips]{graphicx,color}
\fi

\newcommand{\msun}{\ensuremath{\mbox{M}_{\odot}}}

\newcommand{\hii} {\mbox{H$\;$\sc{ii}}}

\newcommand{\heii} {\mbox{He$\;$\sc{ii}}}

\newcommand{\civ} {\mbox{C$\;$\sc{iv}}}

\newcommand{\niv}  {\mbox{N$\;$\sc{iv}}}
\newcommand{\nv}  {\mbox{N$\;$\sc{v}}}

\newcommand{\kms} {\ensuremath{\mbox{km}\;\mbox{s}^{-1}}}

\title[Two new LMC WR stars]
{Two new Wolf-Rayet stars in the LMC}

\author[Ian Howarth \& Nolan Walborn]  
{Ian D.\ Howarth,$^1$\thanks{email: idh@star.ucl.ac.uk}
and Nolan R.\ Walborn$^2$\\
$^1$Dept.\ of Physics and Astronomy, University College London,  Gower
Street, London WC1E 6BT, UK  \\ 
$^2$Space Telescope Science Institute, 3700 San Martin Drive,
Baltimore, MD~21218, USA\thanks{STScI is operated by AURA, Inc., under
NASA Contract NAS 5-26555.}
}

\date{Accepted 2012 July 30.  Received 2012 July 30; in original form 2012 July 5}

\begin{document}

\maketitle

\begin{abstract}
  We report the discovery of two previously unknown WN3 stars in the
  Large Magellanic Cloud.  Both are bright (15th magnitude), isolated,
  and located in regions covered in earlier surveys, although both are
  relatively weak-lined.  We suggest that there may be
  $\mathcal{O}(10)$ remaining undiscovered WNE stars in the LMC.
\end{abstract}

\section{Introduction}
Population~I Wolf-Rayet stars are spectroscopically conspicuous
tracers of the upper end of the initial mass function, with progenitor
masses believed to be in excess of $\sim$25\msun\ \citep{Crowther07}.
In the extragalactic context, Wolf-Rayet (WR) populations are
particularly sensitive to metallicity, such that both the number of
WRs per unit mass (or, equivalently, the WR:O-star ratio), and the
ratio of nitrogen- to carbon-sequence (WN:WC) stars vary considerably
between the Galaxy, Large Magellanic Cloud (LMC), and Small Magellanic
Cloud (SMC).

Because of their strong emission-line spectra, Wolf-Rayets are
relatively easy to identify from very-low-dispersion spectroscopy.
The first systematic search for LMC WRs using objective-prism
spectroscopy yielded 50 objects \citep{Westerlund59}, although many of
these had previously been recorded in the Henry Draper catalogue and
its Extension (e.g., \citealt{Morel88}).  Subsequent objective-prism
searches by \citet{Fehrenbach76}, \citet{Azzopardi79, Azzopardi80},
and \citet{Morgan85, Morgan90} roughly doubled the sample size, with a
handful of other observations, including studies in crowded fields,
leading to the total of 134 entries in the current, `BAT99',
catalogue\footnote{WR status has been disputed for BAT99-4
  \citep{Moffat91}, BAT99-6 \citep{Niemela01}, and BAT99-107
  \citep{Taylor11}. \citet{Massey00} identify Sk$\;-69^\circ194$ as a
  WN3 star with \emph{extremely} weak $\lambda$4686 emission
  ($W_\lambda \simeq -2$\AA), presumably attributable to dilution by
  the companion, although \citet{Foellmi03b} were unable to confirm
  the discovery; and \citet{Evans11} report VFTS~682 as a new, heavily
  reddened WN5h star in 30~Dor.}  \citep{BAT99}.

\begin{figure*}
\center{\includegraphics[scale=0.6,angle=-90]{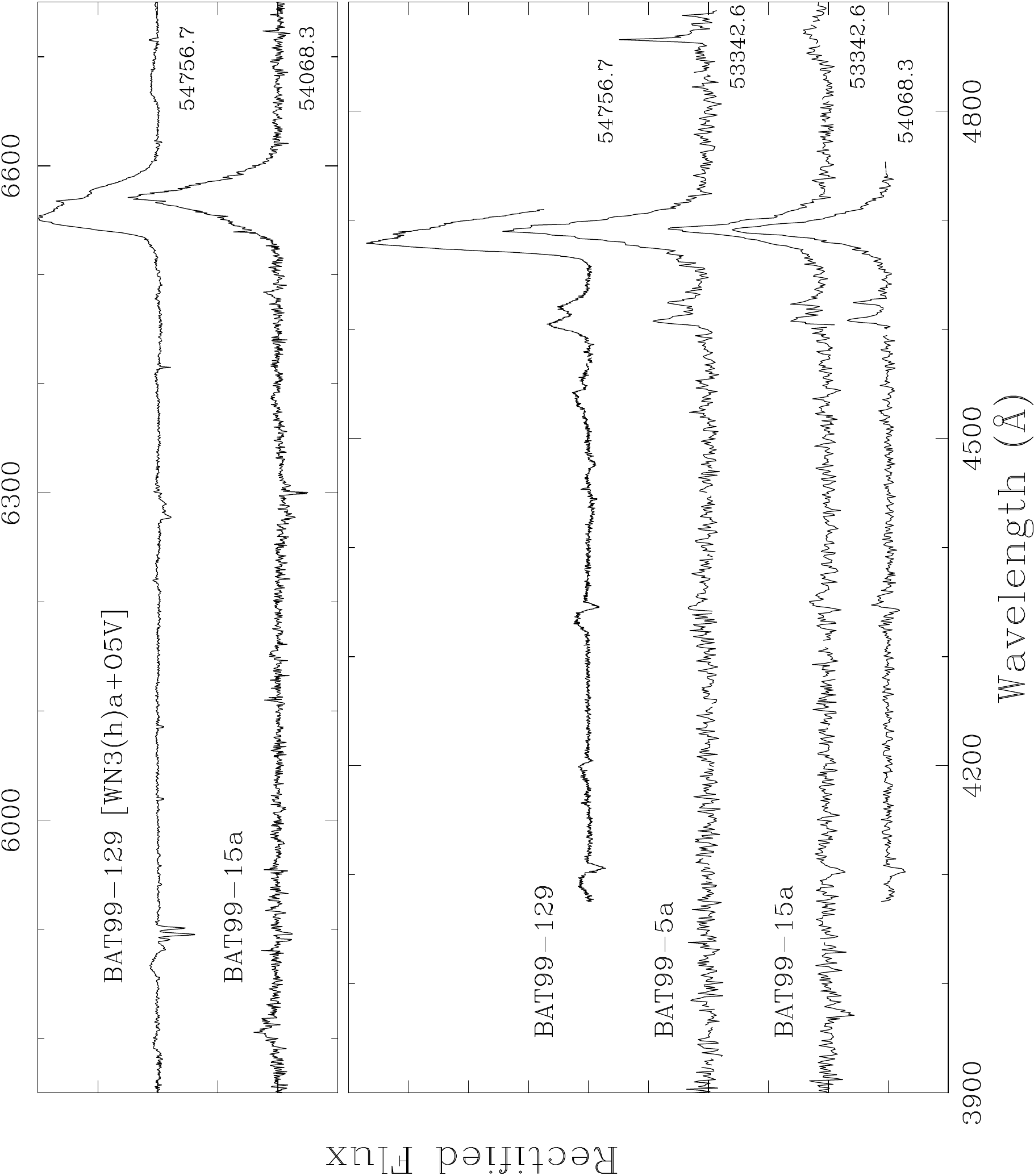}} 
\caption[]{Red (upper panel) and blue spectra of the newly reported LMC WR stars, labelled by
  name and observation date (MJD); earlier observations were obtained
  with the original 2dF system, and later ones with AAOmega.  
An AAOmega spectrum of the large-amplitude SB2 system BAT99-129, classified
WN3ha+O5V
by \citet{Foellmi06}, is shown for comparison.
Ticks on
  the vertical axis are separated by half a continuum
  unit. Rectification is uncertain in the 4800--4900\AA\ region.}
\label{fig_spec}
\end{figure*}

With this tally, the LMC WR population is generally considered to be
known to a substantial degree of completeness.  Any remaining objects
are likely to lie in very crowded regions (although there has been
much progress in this direction in the Hubble Space Telescope era;
e.g., \citealt{Walborn99}), to be heavily reddened (although LMC
reddening is generally small, and intrinsically reddened WC-late stars
are believed to be rare in the LMC), to lie in unexamined outskirts of
the LMC, or to be weak lined.  Here we report the serendipitous
discovery of two bright, isolated WNE stars.  Both are within the area
searched by \citet{Azzopardi80}, but are moderately weak-lined.

\section{Observations}

We have conducted several programmes of spectroscopic observations of
the bright-star content of the LMC ($B \lesssim 15.5$) using the 2dF
and AAOmega instruments on the \mbox{3.9-m} Anglo-Australian
Telescope.  The original 2dF system was a dual-spectrograph,
multi-fibre instrument which allowed up to 400 intermediate-dispersion
spectra to be obtained simultaneously across a $2^\circ$-diameter
field of view \citep{Lewis02}; we used it with 1200B gratings, giving
$R \simeq 2000$ over the wavelength range $\sim$3800--4900\AA.
AAOmega utilises the same fibre-positioner system, but introduces new
spectrographs with many innovations, including dichroic beam-splitting
which allows blue and red spectra to be recorded simultaneously
\citep{Sharp06}.  We used 1700B ($R \simeq 3500, \sim$4100--4750\AA)
and 1000R ($R \simeq 3400, \sim$5650--6800\AA) gratings.

Although poor weather impacted severely on individual programmes, the
powerful multiplexing capability of these instruments has none the
less resulted in the accumulation of spectra of tolerable to good
quality for more than $\sim$3000
%
early-type stars.  An initial examination of this dataset has revealed
two new WR stars.

Neither object has any designation from previous surveys listed in the
CDS \textit{Simbad} database,\footnote{Centre de Donn\'{e}es
  astronomiques de Strasbourg;\newline
  {\tt{http://simbad.u-strasbg.fr/simbad/sim-fcoo}}} so we adopt a
nomenclature based on extending the \citet{BAT99} catalogue.  Basic
stellar parameters are reported in Table~\ref{tab_one}, and the
spectra are shown in Fig.~\ref{fig_spec}.  BAT99-5a is $\sim{5}'$ from
LH~5 and its associated \hii\ regions LHA~120-N83 and N90
\citep{Lucke70,Henize56}, and is $\sim{2.4}'$ from the cluster
NGC~1756, but there is no obvious physical connection with any of
these objects.   BAT99-15a has no nearby LH associations or LHA \hii\ regions.

\subsection{Classification}

Both stars have similar,
early WN types;
although our wavelength coverage doesn't
include the full suite of classification criteria, the lack of
detectable \niv\ $\lambda$4057 coupled with moderately strong \nv\
$\lambda\lambda$4604, 4620 indicates WN3, confirmed by weak \civ\
$\lambda\lambda$5801, 5812 in the case of BAT99-15a.  The
$\lambda\lambda$4340, 4860 emissions in BAT99-15a appear to be
stronger than $\lambda$4541, suggesting hydrogen Balmer emission in
addition to \heii\ Pickering emission (\citealt{Smith96}; these lines are too weak in
BAT99-5a to draw conclusions); it also shows weak absorption at
H$\gamma$--H$\epsilon$ (and possibly at \heii~$\lambda$4541), leading
to a WN3h+abs classification.

[The only plausible alternative
  classifications are as transition Of/WN stars (cf.\
  \citealt{Crowther11}).  Although Fig~\ref{fig_spec} doesn't allow us
  to rule out \emph{weak} P-Cygni absorption at H$\beta$ in BAT99-5a,
  the absence of clear absorption in any other lines favours WN in
  this case, as does the essentially pure-emission H$\beta$ profile in
  the case of BAT99-15a.]
\begin{table}
\caption[]{Basic observational properties;
co-ordinates are from the UCAC-2 catalogue on the ICRS system
\citep{Zacharias04}, and photometry is from \citet{Massey02}.
\nv\ equivalent widths are summed for $\lambda\lambda$4604, 4620.}
\begin{tabular}{lcccccccccccc}
\hline
\multicolumn{13}{c}{BAT99-5a;  WN3} \\
\hline
$\alpha, \delta$ (J2000):&
\multicolumn{6}{r}{4h 55m 7.60s} &
\multicolumn{6}{r}{$-69^\circ\ 12\arcmin\ 31.7\arcsec$} \\
\multicolumn{4}{l}{$B$, $(B-V)$, $(V-R)$:} &
\multicolumn{3}{r}{15.11} &
\multicolumn{3}{r}{$-0.12$} &
\multicolumn{3}{r}{$-0.04$} \\
\multicolumn{4}{l}{$W_\lambda$ (\nv, $\lambda$4686):}&
\multicolumn{3}{r}{$-6.4$\AA} &
\multicolumn{3}{r}{$-40$\AA} \\
\hline
\multicolumn{13}{c}{BAT99-15a; WN3h+abs} \\
\hline
$\alpha, \delta$ (J2000):&
\multicolumn{6}{r}{5h  2m 59.24s} &
\multicolumn{6}{r}{$-69^\circ\ 14\arcmin\  2.3\arcsec$}\\
\multicolumn{4}{l}{$B$, $(B-V)$, $(V-R)$:} &
\multicolumn{3}{r}{15.19} &
\multicolumn{3}{r}{$-0.15$} &
\multicolumn{3}{r}{$-0.20$} \\
\multicolumn{4}{l}{$W_\lambda$ (\nv, $\lambda$4686, $\lambda$6560):}&
\multicolumn{3}{r}{$-4.0$\AA} &
\multicolumn{3}{r}{$-29$\AA} &
\multicolumn{3}{r}{$-40$\AA} \\
\hline
\end{tabular}
\label{tab_one}
\end{table}

\subsection{Intrinsic line strength}

The leptokurtic (`triangular') shape of the \heii~$\lambda$4686
emission in both stars suggests that their spectra are weak-lined
intrinsically (rather than reflecting substantially diluted stronger
emission; e.g., \citealt{Marchenko04}).  

There are 12 stars in the BAT99 catalogue that
have single-star WN3 classifications according to \citet{Foellmi03b};
these have $V$
magnitudes in the range 14.67:16.98 and $(B-V)$ colours $-0.28$:+0.31
according to the compilation by \citet{Bonanos09}.  Assuming
$(B-V)_0 = -0.30$, $R_V = 3.1$ then leads to a range in $V_0$ of
14.4--15.7 for this sample (mean $15.1\pm{0.4}$~s.d.).  With the same
assumptions we find $V_0 =
14.7$, 14.9 for BAT99-5a, 15a, respectively, from the data in
Table~\ref{tab_one}.\footnote{Photometry from \citet{Zaritsky04} gives
  $V_0 = 13.9$, 14.6, but the result for BAT99-5A is compromised by a
  rather large formal error on the $B$-band measurement.}  The absolute
magnitudes are therefore consistent with the known LMC WN3 single-star
population, and with the view that the weak line strengths largely
arise intrinsically, rather than through dilution.

If we nevertheless suppose that the differences in equivalent widths
between 5a and 15a are due solely to modest dilution by an early-type
companion in the latter case, then such a companion would have to
contribute $\sim$25\%\ of the continuum flux; that is, it would have
$B \simeq 16.8$, corresponding to an early-B main-sequence star (with
$V_0 \simeq 15.2$ for an isolated WR primary).  A mix of the spectra
of BAT99-5a and HD~144470 (B1$\;$V) in a continuum ratio of 3:1 does
produce a reasonable match to the observed blue-region spectrum of
BAT99-15a, although the presence of absorption lines by itself does
not necessarily imply an OB binary companion for BAT99-15a.  

[We observed this target at two epochs, two years apart, and the data
show no evidence for variability or for radial-velocity shifts,
whether in absolute terms ($\Delta{V} = +3 \pm 17$~\kms) or
differentially between emission and absorption lines
($\Delta(\Delta{V}) = +1 \pm 42$~\kms).]

\section{Discussion}

\citet{Morgan90} asserted that ``there are probably no undetected WR
stars in the LMC in the $m_v$ range 17--19 except perhaps in obscured
or crowded regions'', and there has been little to challenge that view
in the intervening time.  However, subsequent serendipitous
discoveries (\citealt{Massey00}; \citealt{Evans11}; this paper) hint at the possibility
of significant numbers of brighter undiscovered objects, particularly
among relatively weak-lined WNE stars. (WC stars, with their typically
stronger emission lines, are less likely to have been overlooked.)

Our reasonably well-defined but essentially random sample allows us to
examine this issue.  
By merging results from photometric surveys by
\citet{Zaritsky04}, \citet{Udalski00}, and \citet{Massey02}, we
estimate that there are $\sim$50,000 stars in the LMC brighter than
$B=15.5$.
We have obtained spectra of sufficient quality to identify WR-type
spectra for $\sim$3500 targets, finding two new WNE stars;
together, these numbers 
suggest that there may be as many as $\sim$25 similar objects yet to
be discovered.  Alternatively, we have `rediscovered' 25 known WRs in
our wider dataset, or $\sim$20\%\ of the known population; if we have
discovered a similar proportion of an unknown population, this
instead indicates perhaps $\sim$9 WNEs as yet undiscovered.  Of
course, these estimates are little more than informed guesses given
the small-number statistics, but none the less suggest that perhaps a
dozen or so weak-lined WNE stars, or $\sim$20\%\ of the WNE
population, may remain to be discovered in the LMC.  
If this conjecture proves correct, the dominance of WNE over WNL stars
(and of
WN over WC types) in this metal-poor galaxy   is only strengthened.

\section*{Acknowledgements}

We thank Mike Read (ROE) for supplying objective-prism images, and
Sarah Brough (AAO) for correspondence on 2dF fibre placements, which
together effectively eliminated our initial concerns about potential
misidentification of the targets reported here.  Paul Crowther provided
several very helpful comments, and Rob Sharp (AAO) gave stalwart
support throughout the planning, execution, and reduction of the
observations.


\footnotesize{
\bibliographystyle{mn2e}
\bibliography{IDH}
}

\end{document}